\documentclass{aa}
\usepackage{graphicx}
\usepackage{txfonts}

\begin{document}

\title{From B[e] to A[e]}
\subtitle{On the peculiar variations of the SMC supergiant LHA\,115-S\,23 (AzV\,172)}

\author{M. Kraus\inst{1}, M. Borges Fernandes\inst{2,3},  
  J. Kub\'at\inst{1},
  \and
  F.X. de Ara\'ujo\inst{4}
  }
                                                                                
\institute{Astronomick\'y \'ustav, Akademie v\v{e}d \v{C}esk\'e republiky, Fri\v{c}ova 298, 251\,65 Ond\v{r}ejov, Czech Republic\\
\email{kraus@sunstel.asu.cas.cz; kubat@sunstel.asu.cas.cz}
 \and
Royal Observatory of Belgium, Ringlaan 3, B-1180 Brussels, Belgium
 \and
UMR 6525 H. Fizeau, Univ. Nice Sophia Antipolis, CNRS, Observatoire de
la C\^{o}te d'Azur, Av. Copernic, F-06130 Grasse, France\\
\email{Marcelo.Borges@obs-azur.fr}
 \and
   Observat\'orio Nacional, Rua General Jos\'e Cristino 77, 20921-400 S\~ao Cristov\~ao, Rio de Janeiro, Brazil\\
 \email{araujo@on.br}
      }
                                                                                
\date{Received; accepted}

\abstract
%Context
{Optical observations from 1989 of the Small Magellanic Cloud (SMC) B[e] 
supergiant star LHA\,115-S\,23 (in short: S\,23) revealed the presence of 
photospheric He{\sc i} absorption lines, classifying S\,23 as a B8 supergiant. 
In our high-resolution optical spectra from 2000, however, we could not 
identify any He{\sc i} line. Instead, the spectral appearance of S\,23 is more
consistent with the classification as an A1 supergiant, maintaining the 
so-called B[e] phenomenon.}
%Aims
{The observed changes in spectral behaviour of S\,23 lead to different
spectral classifications at different observing epochs. The aim of this 
research is, therefore, to find and discuss possible scenarios that might cause 
a disappearance of the photospheric He{\sc i} absorption lines within a period 
of only 11 years.}
%Methods
{From our high-resolution optical spectra, we perform a detailed investigation 
of the different spectral appearances
of S\,23 based on modern and revised classification schemes. In particular, we 
derive the contributions caused by the interstellar as well as the 
circumstellar extinction self-consistently. The latter is due to a partly 
optically thick wind. We further determine the projected rotational velocities 
of S\,23 in the two epochs of spectroscopic observations. 
}
%Results
{Based on its spectral appearance in 2000, we classify S\,23 
as A1\,Ib star with an effective temperature of about 9000\,K.
This classification is supported by the additional analysis of the photometric 
$UBV$ data. An interstellar extinction value of $E(B-V) \simeq 0.03$ is
derived. This is considerably lower than the previously published value,
which means that, if the circumstellar extinction due to the stellar wind is 
neglected, the interstellar extinction, and hence the luminosity of the star, 
are overestimated. We further derive a rotation velocity of $\varv\sin 
i \simeq 150$\,km\,s$^{-1}$, which means that S\,23 is rotating with about 75\%
of its critical speed. The object S\,23 is thus the fourth B[e] supergiant with confirmed
high projected rotational velocity. The most striking result is the apparent 
cooling of S\,23 by more than 1500\,K with a simultaneous increase of its 
rotation speed by about 35\% within only 11\,years. Since such a behaviour is 
excluded by stellar evolution theories, we discuss possible scenarios for the
observed peculiar variations in S\,23.  
}
%Conclusions
{}


\maketitle

\section{Introduction}

B[e] stars belong to one of the puzzling phenomena in modern astrophysics. 
While many of their members have been known for decades, the number of new 
detections of B[e] stars is growing tremendously (see, e.g., Kraus \& 
Miroshnichenko \cite{KM}). The classification by Lamers et al. (\cite{Lamers98})
of B[e] stars according to their evolutionary phase made clear that B[e] stars 
occur in the pre- as well as in the post-main sequence phase, and the most 
popular class is formed by the B[e] supergiants with confirmed members 
predominantly in the Magellanic Clouds (see the recent review by Zickgraf 
\cite{Zickgraf2006}).

The derivation of stellar parameters of B[e] stars is complicated because in 
most cases the stellar photospheres are hidden inside the dense and optically
thick winds. Thus, the spectra are contaminated with wind continuum emission in 
the form of free-free and free-bound emission as well as with an enormous 
amount of emission lines that are generated in the circumstellar wind material, 
which is usually non-spherically symmetric. Evidence for non-sphericity thereby 
comes from polarimetric observations, e.g., by Magalh\~aes (\cite{Magalhaes}), 
Magalh\~aes et al. (\cite{Magalhaesetal}), and Melgarejo et al. 
(\cite{Melgarejo}), in rare cases even from optical imaging with the 
Hubble Space Telescope, as was the case for the galactic B[e] star Hen\,2-90 
(Sahai et al. \cite{Sahai02}; Kraus et al. \cite{Kraus05}), and from 
interferometric observations with, e.g, the VLTI/AMBER and VLTI/MIDI 
instruments (e.g., Domiciano de Souza et al. \cite{Armando}). An observed
strong infrared (IR) excess emission due to warm and hot circumstellar dust 
(e.g., Zickgraf et al. \cite{Zick85}, \cite{Zickgraf86}, \cite{Zickgraf89}, 
\cite{Zick92}; Gummersbach et al. \cite{Gummersbach}; Borges Fernandes et al.
\cite{Marcelo}) indicates the presence of a high-density disk. This disk must
further be the location of the neutral material like O{\sc i} (Kraus et al. 
\cite{Kraus07}) and the molecular gas, expressed by CO first-overtone bands 
(McGregor et al. \cite{McGregorA}; \cite{McGregorB}; \cite{McGregor}; Morris et 
al. \cite{Morris}; Kraus et al. \cite{CO}) and TiO band emission (Zickgraf et 
al. \cite{Zickgraf89}). The proposed scenario for B[e] supergiants thus 
suggests that their circumstellar material consists of a disk-like structure 
formed by a high density equatorial wind with low outflow velocity, and a 
normal B-type line-driven wind of high velocity and much lower density in polar 
directions (Zickgraf et al. \cite{Zick85}; Zickgraf \cite{Zickgraf2006}). A 
possible mechanism for the formation of outflowing high density disks might be 
rapid rotation of the central star (see, e.g., Bjorkman \& Cassinelli 
\cite{Bjorkman}; Cur\'e \cite{Cure04}; Cur\'e et al. \cite{Cure05}). In 
addition, if the star is rotating rapidly, the wind material in the equatorial 
plane recombines in hydrogen in the vicinity of the stellar surface, as has 
recently been shown by Kraus (\cite{Kraus06}). And for at least three B[e] 
supergiants projected rotational velocities at a significant fraction of their 
critical speed have indeed been found (Gummersbach et al. \cite{Gummersbach}; 
Zickgraf \cite{Zickgraf2000}; \cite{Zickgraf2006}), hinting to the possible 
connection between the presence of a circumstellar disk and stellar rotation.

For the Small Magellanic Cloud (SMC) star \object{LHA 115-S 23} (AzV 172, 2MASS
J00555380-7208596, in the following refered to as S\,23) that is investigated in
this paper, no rotational velocity has been derived yet. In addition, not much is 
known about the orientation of the system, although it has been suggested to be
seen more or less edge-on, based on the low expansion velocity derived from the
Balmer lines (Zickgraf et al. \cite{Zick92}, referred to as ZSW in the 
following). This object was first classified by Azzopardi \& Vigneau 
(\cite{AV}) as a B8 supergiant, and the detection of several photospheric 
He{\sc i} absorption lines in the optical mid-resolution spectrograms of ZSW is 
in agreement with the classification of S\,23 as a late B-type supergiant. 
These authors further fixed the interstellar extinction at $E(B-V) = 0.10$, 
derived an effective temperature of $T_{\rm eff} = 11\,000$\,K, and a surface 
gravity of $\log g = 2.0$. This led to $M_{\rm bol} = -6.4, \log L_*/L_{\odot} 
= 4.46, R_* = 45\,R_{\odot}$, and the assignment of the luminosity class Ib. 
The spectra of ZSW showed further strong Balmer line emission, [O{\sc i}] 
emission, as well as many Fe{\sc ii} and [Fe{\sc ii}] emission lines. In 
combination with the pronounced infrared excess in the photometric data, which 
can be ascribed to warm or hot circumstellar dust, ZSW assigned S\,23 to the 
group of B[e] supergiants.

About eleven years later, during our observing mission of Magellanic Cloud B[e] 
supergiants, we observed S\,23 using a high-resolution optical spectrograph. 
Our spectra confirm the presence of the strong Balmer emission lines as well as 
of the [O{\sc i}], [Fe{\sc ii}], and Fe{\sc ii} emission lines. But 
interestingly, even with a more than 10 times higher resolution, our spectra do 
not display any He{\sc i} line either in absorption nor in emission. The 
absence of the  He{\sc i} lines speaks more in favour of a cooler central 
object than a late B supergiant. This paper is therefore aimed to investigate
in detail the spectral appearances of S\,23 at the two epochs of observations
and to discuss possible scenarios that might cause the observed peculiar 
variations of S\,23.

The paper is structured as follows: our observations are described in 
Sect.\,\ref{obs}. Then, we perform the spectral classification of S\,23 for the 
two different observing epochs, using modern, improved classification schemes 
(Sect.\,\ref{specclas}) and determine the contributions of both, interstellar
and circumstellar extinctions (Sect.\,\ref{extinc}). From the detected 
photospheric Mg{\sc ii} absorption lines, we next derive the projected
rotation velocity of S\,23 at the two epochs of observations 
(Sect.\,\ref{rotation}). In Sect.\,\ref{discus}, we discuss possible scenarios
that might explain the observed peculiarities, before we summarise our major
results in Sect.\,\ref{concl}.

\section{Observations}\label{obs}

\begin{figure}[t!]
\resizebox{\hsize}{!}{\includegraphics{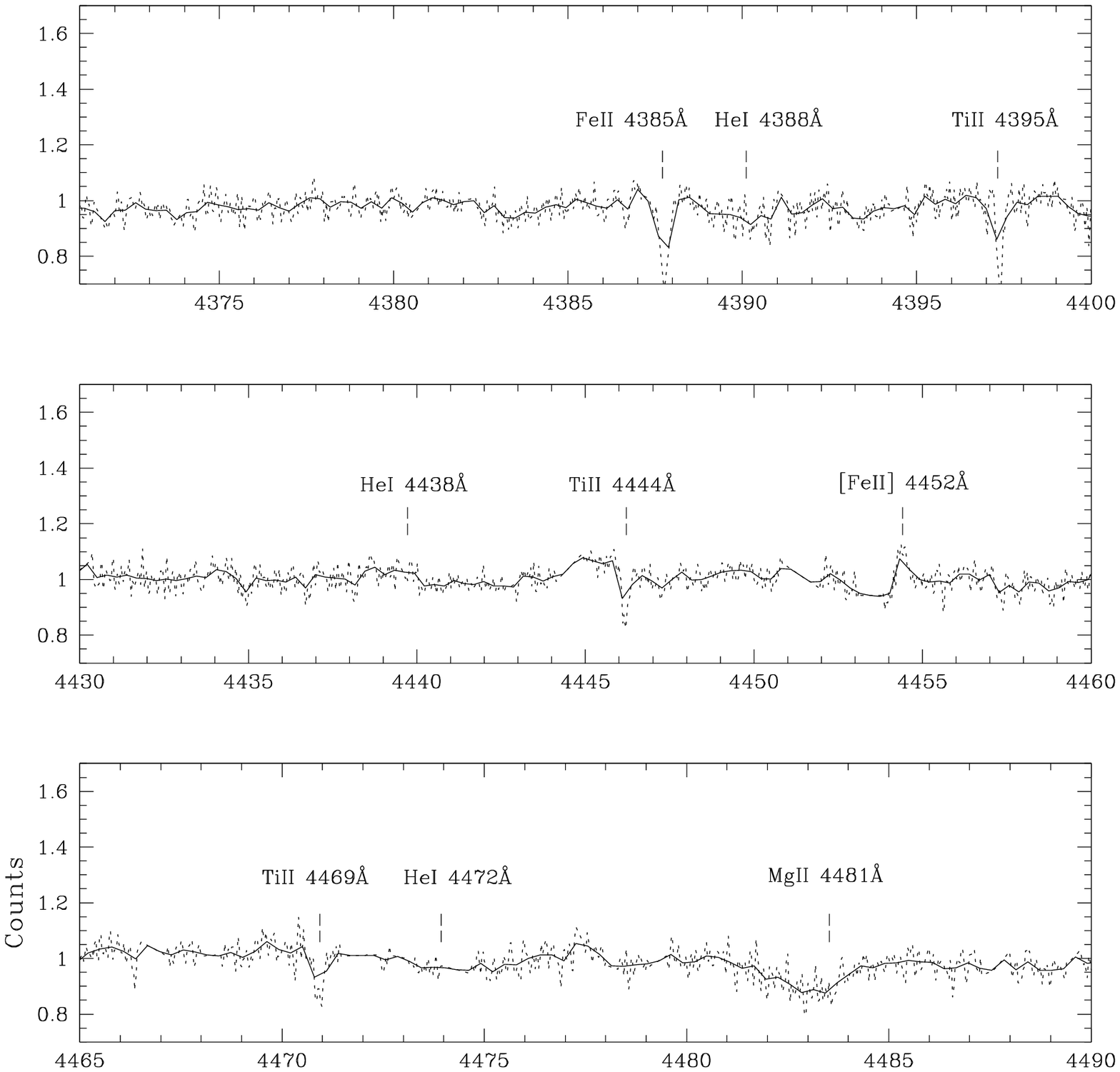}}
\resizebox{\hsize}{!}{\includegraphics{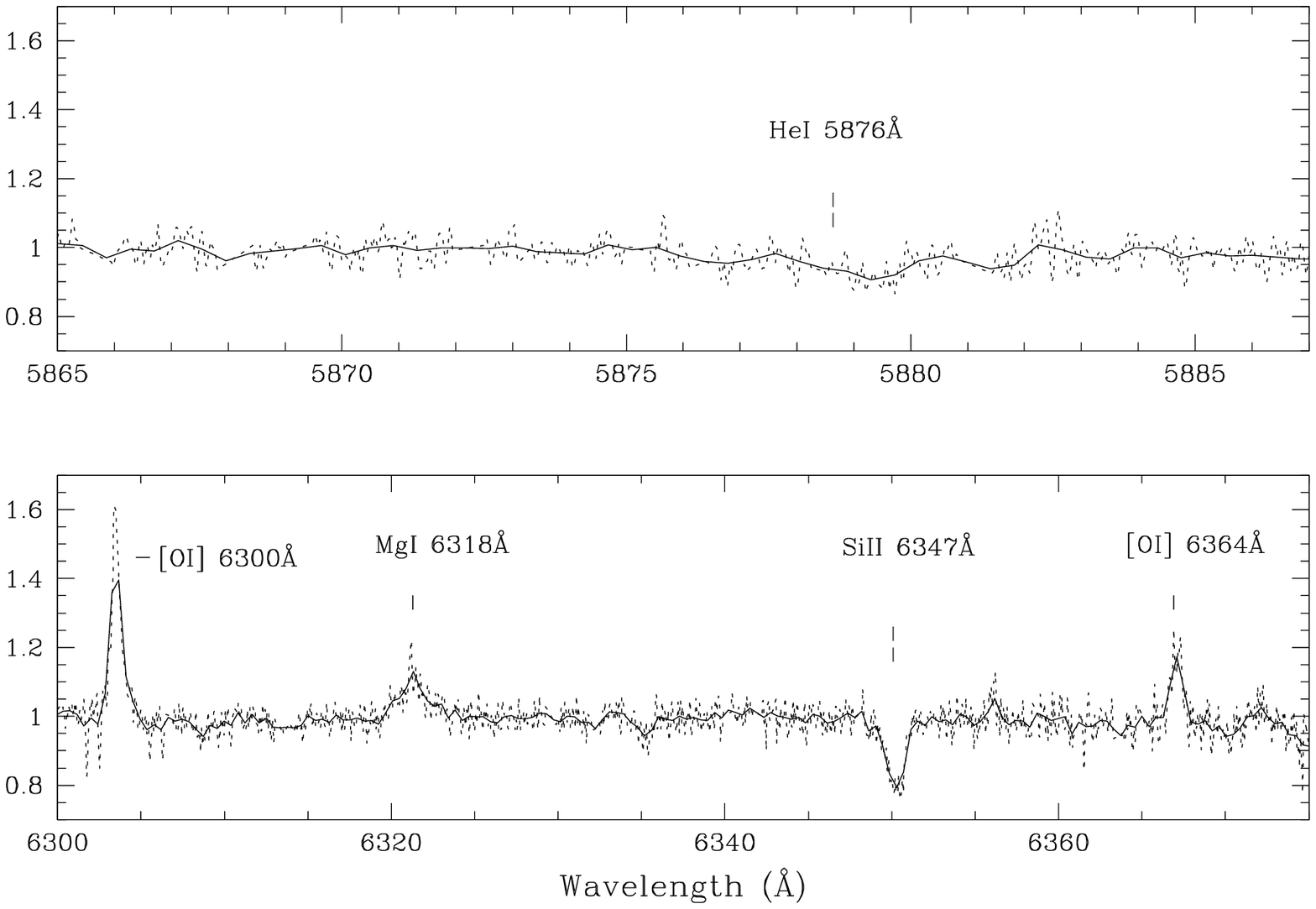}}
\caption{Parts of the rectified FEROS spectrum (dashed lines) showing the
regions around several expected He{\sc i} lines. For better visualization, we
convolved the FEROS data to a resolution of $R = 4900$ as was used by ZSW
(solid lines). 
}
\label{first}
\end{figure}

We obtained the optical spectra with the 1.52-m telescope at the European 
Southern Observatory in La Silla 
(Chile) using the high resolution Fiber-fed Extended Range Optical Spectrograph 
(FEROS).
FEROS is a bench-mounted Echelle spectrograph with fibers, which covers a sky
area of 2$\arcsec$ of diameter, with a wavelength coverage from 3600\,\AA \ to 
9200\,\AA \ and a spectral resolution of R = 55\,000 (in the region around 6000 
\AA). We adopted its complete automatic on-line reduction. 
The high-resolution spectrum of S\,23 was taken on October 14, 2000, with an 
exposure time of 3600 seconds. The S/N ratio in the 5500\,\AA \ region is 
approximately 40. Since within the same night we were able to take two 
consecutive spectra of the star, which do not show significant differences, 
we added them up for a better S/N ratio achievment.

In Fig.\,\ref{first}, we display parts of the rectified high-resolution FEROS 
spectra. The first three panels show the regions around the positions of 
several He{\sc i} lines, which were identified by ZSW in their 1989 
low-resolution (i.e. $R=4900$) observations (see Fig.\,4 of ZSW), with the 
He{\sc i} $\lambda$4472 line being the strongest one with almost equal 
equivalent width to the adjacent Mg{\sc ii} $\lambda$4481 
line. This He{\sc i} line is clearly absent in our FEROS spectrum (or at least
below the detection limit) as well as the other two He{\sc i} lines at 
$\lambda\lambda$4388, 4438, whose detection was previously reported. 
The fourth panel of Fig.\,\ref{first} displays the region around the 
He{\sc i} $\lambda$5876 line. If He{\sc i} is present in a stellar spectrum, 
then this line is usually quite strong. But obviously, it is also absent from 
our FEROS spectrum. We checked the complete FEROS spectrum, but we could not 
identify a single He{\sc i} line.  

Due to our rather low S/N ratio and the fact that 
the photospheric He{\sc i} absorption lines are expected to be broad and 
shallow, the high-resolution spectrum might not be able to display these lines 
properly. Instead, for these lines a (much) lower spectral resolution might be 
advantageous as has been shown by Verschueren (\cite{Versch}). We therefore 
convolved our data to the same resolution as the ZSW spectrum and overplotted 
the low-resolution spectrum in Fig.\,\ref{first}. However, even with the 
low-resolution data, the He{\sc i} lines do not show up.

\begin{figure}[t!]
\resizebox{\hsize}{!}{\includegraphics{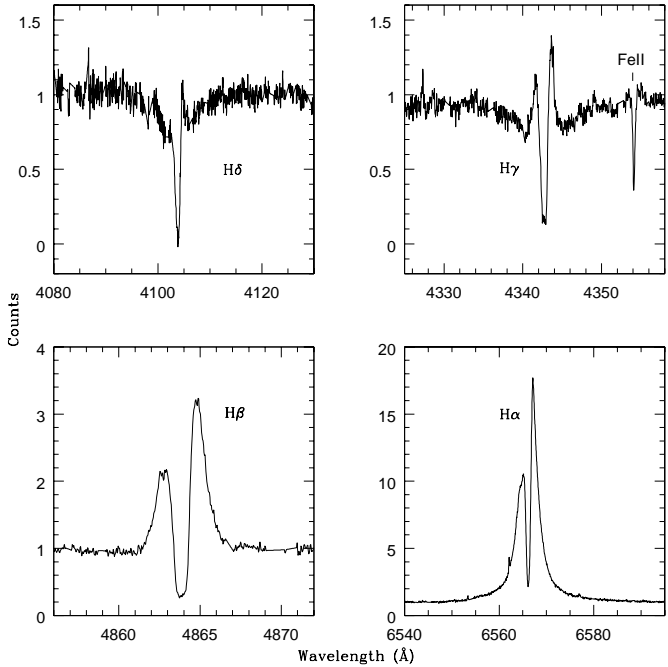}}
\caption{Balmer lines in the spectrum of S\,23.}
\label{Balmer}
\end{figure}

While other known B[e] supergiants do not display even a single 
photospheric absorption line in their optical spectra due to their high 
density and therefore optically thick winds, S\,23 displays at least one 
clearly detectable line, namely the Mg{\sc ii} doublet at $\lambda$4481. And 
also the higher Balmer lines (H$\delta$, H$\gamma$, see Fig.\,\ref{Balmer}), 
as already mentioned by ZSW, show indications for a photospheric absorption 
component on which P\,Cygni profiles seem to be superimposed. Some other lines 
in our spectra also display absorption components, like Fe{\sc ii}, 
Ti{\sc ii}, and Si{\sc ii} (see Fig.\,\ref{first}). These absorption components
are, however, much narrower than the Mg{\sc ii} line, indicating that they
are circumstellar in origin. 

\begin{table}[t !]
\caption{Mean heliocentric radial velocities derived from different ions
compared to those derived by ZSW.} 
\label{radvelo}
\begin{tabular}{lcr}
    \hline
    \hline
Ion & $\varv_{\rm rad}$ & $\varv_{\rm rad}^{\rm ZSW}$ \\ 
 & (km\,s$^{-1}$) & (km\,s$^{-1}$) \\
    \hline
$[$Fe\,{\sc ii}] & 156.9 & 141.0 \\
$[$O\,{\sc i}] & 156.1 & 164.0 \\
Si\,{\sc ii} & 155.7 & 150.0 \\
Fe\,{\sc ii} & 159.9 & 144.0 \\
Cr\,{\sc ii} & 161.1 & --- \\
Balmer lines & 152.8 & 143.0 \\
H$\alpha$$^{\rm ca}$ & 150.9 & 126.0 \\
H$\alpha$$^{\rm bp}$ & 110.0 & 70.0 \\
H$\alpha$$^{\rm rp}$ & 197.0 & 210.0 \\
    \hline
  \end{tabular}
\end{table}
                                                                                
We measure the heliocentric radial velocity for several lines in our spectra: 
from the narrow pure emission lines of the forbidden transitions (like 
[O{\sc i}], [Fe{\sc ii}], see Fig.\,\ref{first}), from the narrow absorption 
components observed for most of the permitted transitions (like Fe{\sc ii}, and 
two lines of Si{\sc ii} and Cr{\sc ii}), and from the narrow absorption 
components of the Balmer lines\footnote{For the
Balmer lines, we list the mean radial velocity derived from the central
absorption component for 4 lines (H$\beta$, H$\gamma$, H$\delta$ and
H$\epsilon$ - H$\beta$ and H$\epsilon$ were not observed by ZSW), while for
H$\alpha$ the radial velocity derived for the central absorption ($^{\rm ca}$),
the red peak ($^{\rm rp}$), and the blue peak ($^{\rm bp}$) are given.}.
For each element we derive mean values, which are listed in 
Table\,\ref{radvelo}. For comparison, we also included in this table the
values obtained by ZSW. While our metal lines all deliver values in the range 
$155\ldots 160$\,km\,s$^{-1}$, the values of ZSW indicate a much larger 
scatter between the radial velocities of the forbidden and the pemitted lines,
which had been interpreted by these authors as a slow outflow component. 
In fact, the narrow absorption components of the permitted lines, often 
overlaid on broader emission components, are not of pure atmospheric origin.
Instead, they are generated due to absorption by the circumstellar material.
The same holds also for the narrow absorption components seen in the Balmer
lines. For a proper radial velocity determination we therefore rely on  
forbidden lines, which display narrow but symmetrical line profiles.
From these lines, we thus obtain a mean radial velocity of $\varv_{\rm rad} = 
156.5\pm 1.0$\,km\,s$^{-1}$, which we will use here. 
It is slightly lower than the value found by ZSW from their [O{\sc i}]
lines, but this might be a spectral resolution effect rather than a stellar
effect.

The largest difference between the ZSW and our data is reached by the two
emission peaks of the H$\alpha$ line. Here, a clear shift in velocities is
present. This shift is about 40\,km\,s$^{-1}$ for the blue peak, while it is
only 10--15\,km\,s$^{-1}$ for the red peak. We tested the influence of the 
different resolutions by smearing our H$\alpha$ line to the same resolution as 
ZSW. The measured radial velocities are in much better agreement with the 
ZSW values, so that we can conclude that the differences in the appearance of 
the H$\alpha$ line in ZSW and in our spectrum are most probably caused by the 
different spectral resolutions.

\section{Spectral classification of S\,23}\label{specclas}

\subsection{Temperature classification}\label{tempclass}

The determination of the spectral type and luminosity class of stars with 
metallicities different from the galactic one has been found to be a difficult 
task in the past. The reason for this is the fact that with decreasing 
metallicity the metal lines in the spectra become very weak and hard to  
detect. The proper determination of the spectral type is rather complicated, 
especially for B-type stars which do not or only very weakly show He{\sc ii}
lines in their spectra. For these objects, the classification has to be 
performed with the help of  
the metal lines and is, for galactic objects, normally based on the line
strengths of silicon and magnesium, usually in comparison to helium. As shown, 
e.g., by Lennon et al. (\cite{Lennon93}) for galactic B supergiants, the 
equivalent widths of many spectral lines show a typical behaviour with spectral 
type, i.e. effective temperature. And later on, Lennon (\cite{Lennon97}) 
suggested determining the spectral classes of B-type supergiants in the SMC 
according to line ratios in such a way that the same trend in line ratios 
with spectral type is achieved as for solar metallicity (Lennon et al. 
\cite{Lennon92}). His classification scheme (see Table\,1 of Lennon, 
\cite{Lennon97}), which has been extended to G-type stars by Evans et al. 
(\cite{Evans}), is used by many authors to classify B-type supergiants in many 
galaxies with abundances similar to the SMC value (e.g., Evans \& Howarth 
\cite{EvansHowarth}; Bresolin et al. \cite{Bresolin06}; Bresolin et al. 
\cite{Bresolin07}).

The spectral type of S\,23 has been determined by Azzopardi \& Vigneau 
(\cite{AV}) as B8, and later on, ZSW, derived an effective temperature of 
about 11\,000\,K based on comparison of the observed 
spectral energy distribution to Kurucz (\cite{Kurucz}) model atmospheres.
Inspection of their mid-resolution spectrum indicates the presence
of the He{\sc i} $\lambda$4472 and Mg{\sc ii} $\lambda$4481 lines, with 
the Mg{\sc ii} line being slightly stronger than the He{\sc i} line.
Lennon et al. (\cite{Lennon93}) already mentioned the sharp decrease in 
He{\sc i} equivalent width from early to late B-type supergiants, and the 
simultaneous strong increase in Mg{\sc ii} $\lambda$4481 equivalent width.
The relative strength of these two lines is therefore one of the main 
classification characteristics of especially late B-type supergiants 
(e.g., Bresolin et al. \cite{Bresolin06}; Lennon \cite{Lennon97}).
According to the classification scheme of Lennon (\cite{Lennon97}), the line
ratio as inferred from the ZSW spectrum would thus suggest a spectral type 
between B8 and B9 of S\,23 in the year 1989.

\begin{figure}[t!]
\resizebox{\hsize}{!}{\includegraphics{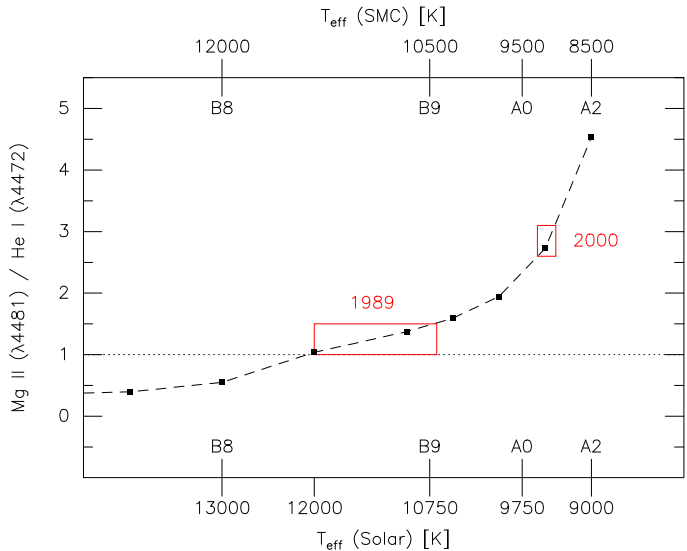}}
\caption{Effective temperature dependence of the line ratio Mg{\sc ii}
$\lambda$4481 over He{\sc i} $\lambda$4472 calculated from model stellar
atmospheres with $\log g = 2.5$ at solar abundances. 
The behaviour between spectral type
and effective temperature taken from Evans \& Howarth (\cite{EvansHowarth}) is
indicated in the bottom label for solar abundance, and in the top label for SMC
abundance. The two boxes indicate the observed ratios with their errors for
the ZSW spectrum taken in 1989 and our spectrum taken in 2000, respectively.}
\label{ratio}
\end{figure}

Our own high-resolution spectra display only the Mg{\sc ii} $\lambda$4481 line
while the adjacent He{\sc i} $\lambda$4472 line is definitely below the
detection limit. This would speak more in favour of a star with spectral type 
A0 or later. To quantify these classifications and to constrain the possible
range in effective temperature at the two observing epochs, it is necessary to 
calculate theoretical line ratios of Mg{\sc ii} $\lambda$4481 over He{\sc i} 
$\lambda$4472. This is done by using the code SYNSPEC (see Hubeny \& Lanz 
\cite{Hubeny}) to compute line identification tables based on Kurucz
(\cite{Kurucz}) model atmospheres in local thermodynamical equilibrium (LTE)
and Kurucz (\cite{Kurucz93}) line lists. These calculations have been performed 
for effective temperatures ranging from 8\,000\,K up to 15\,000\,K and solar 
abundances. With these line identification tables that contain information 
about the expected line equivalent widths, we then calculated the Mg{\sc ii} 
$\lambda$4481 over He{\sc i} $\lambda$4472 line ratios. The results are shown 
as the dashed line in Fig.\,\ref{ratio}. To convert the effective temperature 
scale into a spectral type, we made use of the recent classification of Evans 
\& Howarth (\cite{EvansHowarth}), who redefined effective temperatures for 
galactic (indicated along the bottom axis of Fig.\,\ref{ratio}) and SMC 
supergiants (indicated along the top axis). In addition, we follow the approach 
of Lennon (\cite{Lennon97}), who suggested that the same line ratio mirrors the 
same spectral type. We find that the transition from a line ratio smaller than 
one to a ratio larger than one happens at an effective temperature of about 
12\,000\,K for solar abundances, which corresponds to a spectral type between 
B8 and B9, in agreement with what has been claimed by Lennon (\cite{Lennon97}).

\begin{table*}[t!]
\caption{Photometric data of S\,23 (in mag) collected from the literature
and sorted by their year(s) of observation.}
\label{photo}
  \begin{tabular}{lcccccccl}
    \hline
    \hline
 Year &  $U$ & $B$ & $V$ & $R$ & $J$ & $H$ & $K/K_{s}$ & References \\
    \hline
 1972-74 & 13.27 & 13.43 & 13.37 & -- & -- & -- & -- & Azzopardi \& Vigneau (\cite{AV})  \\
 1988    & -- & -- & -- & -- & 13.04 & 12.79 & 12.16 & Zickgraf et al. (\cite{Zick92}) \\
 1998    & -- & -- & -- & -- & 13.02 & 12.76 & 12.15 & Skrutskie et al. (\cite{2MASS}; 2MASS) \\
 1999    & 13.11 & 13.38 & 13.31 & 13.19 & -- & -- & -- & Massey (\cite{Massey}) \\
    \hline
  \end{tabular}
\end{table*}

Concerning the measured line ratios of S\,23 we could derive a value of
$1.25\pm 0.25$ from the low-resolution spectrum of ZSW\footnote{This value has
been derived by hand from the published spectrum because the digital spectra
are not available anymore (Zickgraf, private communication).}. This value, with
its error, is included in Fig.\,\ref{ratio}. Its location in the diagram 
indicates that in 1989 the spectrum of S\,23 was that of a B8.5--B9 star with 
effective temperature $T_{\rm eff} \simeq (11\,000\pm 500)$\,K. This 
classification is in fair agreement with the one derived by ZSW.

From our own spectrum taken in 2000, we derive a line ratio of $2.85\pm 0.25$.
This ratio has to be considered as a lower limit since we used, for the
equivalent width of the He{\sc i} $\lambda$4472 line, the value we consider
the detection limit. With this line ratio, we therefore determine an upper
limit to the effective temperature that is mirrored by this line ratio.
We find, for S\,23, a spectral type later than A0 and an effective temperature
$T_{\rm eff} \la 9\,500$\,K. If the spectral appearance of S\,23 at the two
observing epochs is indeed determined by the temperature of the central star,
then the change in line ratio suggests that the star must have cooled by at
least 1500\,K within a period of 11 years only, a scenario that is 
definitely excluded by stellar evolution theory (see Sect.\,\ref{stelevol}).
In the following, we will therefore refer to this behaviour as an {\it 
apparent} cooling of the star.

\subsection{Luminosity class}\label{lumclass}

For the SMC, great effort has been undertaken for a proper luminosity
classification of the supergiant population. In principle, several methods for
the determination of the stellar luminosity exist and have been successfully
applied in the literature. For
O-type to early B-type supergiants, the He{\sc i}/He{\sc ii} line ratios are
good and, especially, metallicity-independent luminosity indicators (e.g., 
Walborn \cite{Walborn}). For B and later type supergiants, the H$\gamma$ 
equivalent width is an ideal and metallicity-independent tool for the 
luminosity class
determination, as shown by Azzopardi (\cite{Azzopardi}), while the use of metal
lines as suggested by Massey et al. (\cite{Massey95}) delivers unreliable
results when applied to stars in low-metal content galaxies like the SMC (see
the discussion in Lennon \cite{Lennon97}).

For S\,23, the luminosity class determination is complicated due to fact
that its H$\gamma$ line shows an additional emission component (see
Fig.\,\ref{Balmer}). In this case, the equivalent width method of Azzopardi
(\cite{Azzopardi}) is not applicable.
                                                                                
ZSW had derived the luminosity class Ib from determining $E(B-V)$ based on the 
photometric $B$ and $V$ data of S\,23 from Azzopardi \& Vigneau (\cite{AV}). 
Using the intrinsic colors for a B8 supergiant from Schmidt-Kaler (\cite{SK}),
they found a value of $E(B-V) = 0.10$. However, as we will show in 
Sect.\,\ref{extinc}, this extinction determination is based on assumptions 
that are not reliable for typical B[e] stars, and consequently, the $E(B-V)$ 
value derived by ZSW {\it overestimates} the real amount of interstellar 
extinction and, therefore, the luminosity of S\,23.
                                                                                
From our FEROS spectrum alone, we are not able to make any estimates about the
stellar luminosity. We therefore searched the literature for complementary 
photometric data. These data, which have usually errors on the order of 
0.01\,mag, are summarised in Table\,\ref{photo}, and the $K_{s}$ band belongs 
to the 2MASS observations. Inspection of this table delivers the following 
important information:

\begin{itemize}
\item The $UBV$ data of S\,23 have changed substantially between the 
observations taken in 1972-74 and those taken in 1999, while the near-IR 
photometry appears constant in the measurements taken in 1988 and in 1998.
\item The $UBV$ data of Azzopardi \& Vigneau (\cite{AV}) were taken
14--16 years {\it before} ZSW took their near-IR data. With the knowledge of
a possible variation/cooling of the central star as discussed above,
the combination of these two datasets for the construction of the spectral
energy distribution is therefore most certainly unreliable.
\item In 1999, the star appears to be more luminous in the $UBV$ bands than
it was during the 1972-74 observing period. 
\item The $UBV $ data of Massey (\cite{Massey}) were taken only one year
earlier than our FEROS spectra, and we use them as complementary data to our 
optical spectra for a tentative luminosity determination. 
\end{itemize}

From the photometry of Massey (\cite{Massey}) we can estimate the luminosity 
class of S\,23 in the year 1999, if we completely neglect the interstellar 
extinction towards S\,23 for the moment. From the distance modulus of $\mu = 
18.93$ (Keller \& Wood \cite{Keller}), we thus find a first-guess absolute 
visual magnitude of $M_{V} = -5.62$, assigning to S\,23 the luminosity class Ib.

\section{Interstellar and circumstellar extinction}\label{extinc}

In the previous section, we determined the luminosity class of S\,23 as Ib. For 
this, we ignored the presence of any interstellar extinction value, knowing
that this delivers only a lower limit to the real stellar luminosity. In this 
section, we therefore intend to derive the values of the interstellar and 
possible circumstellar extinction. This is necessary for a proper dereddening 
of the photometric data and a reliable stellar luminosity determination. 

To derive $E(B-V)$ from the color index of a star, one should not only rely on 
the $B-V$ color index, but also check, if possible, the result by using a 
different color index, e.g. $U-B$. The observed colors from Azzopardi \& 
Vigneau (\cite{AV}) given in Table\,\ref{photo} serve as an example. According 
to the classification of S\,23 as B8\,Ib, ZSW derived a value of $E(B-V) = 0.10$ 
from $(B-V)$. Using as well the color index $(U-B)$, and the intrinsic color
of Johnson (\cite{Johnson}), we find an extinction of $E(B-V) = 0.49$. For the 
calculation of this value, we used the relation $E(B-V) = E(U-B) / 0.77$, which
has been found from observations by Leitherer \& Wolf (\cite{Leitherer}). The 
value found from the $U-B$ color index thus seems to indicate an extinction 
five times higher. 

One might argue that, since the star seems to be somehow variable, the spectral 
type derived from the optical data of ZSW cannot be applied to the $UBV$ data 
taken about 14 years earlier. Therefore, we checked the possible interstellar 
extinction values derived from the more recent set of $UBV$ data from Massey 
(\cite{Massey}), using the spectral classification derived from our optical 
spectra that have a time offset of about one year, only.

From Fig.\,\ref{ratio}, we found the spectral type of S\,23 as A0 or later,
and in the following we will deal with two stellar scenarios, namely an A0\,Ib
star with $T_{\rm eff} = 9500$\,K, and an A1\,Ib star with $T_{\rm eff} = 
9000$\,K. Using the tables with intrinsic colors from Johnson (\cite{Johnson}),
we find the following extinction values: 
\begin{displaymath}
E(B-V) = \left\{\begin{array}{c}
   0.06 \\
   0.18
 \end{array}  \right\} \quad \mathrm{from} \quad
\left\{\begin{array}{c}
(B-V) \\
(U-B)
\end{array}  \right\} \quad \mathrm{for} \quad \mathrm{A0\,Ib}\, ,
\end{displaymath}
and
\begin{displaymath}
E(B-V) = \left\{\begin{array}{c}
   0.04 \\
   0.07
 \end{array}  \right\} \quad \mathrm{from} \quad
\left\{\begin{array}{c}
(B-V) \\
(U-B)
\end{array}  \right\} \quad \mathrm{for} \quad \mathrm{A1\,Ib}\, .
\end{displaymath}
Again, both methods (even considering an error in the $UBV$ band fluxes 
of about 0.01\,mag) deliver results that differ by a factor 2--3. This 
discrepancy in extinction implies that, obviously, the interstellar reddening 
is not the only extinction source acting on the observed $UBV$ data. Instead, 
circumstellar material along the line of sight is causing an extinction of the 
stellar flux as well. 

Possible circumstellar extinction sources might be, on the one hand, 
circumstellar dust with a wavelength dependent extinction behaviour different 
from the the global interstellar one. On the other hand, since B[e] stars are known 
to have high mass loss rates, the circumstellar extinction might be caused by 
the high density wind that drops partly optically thick at optical and/or UV 
wavelengths. 

\begin{figure*}[t!]
\resizebox{\hsize}{!}{\includegraphics{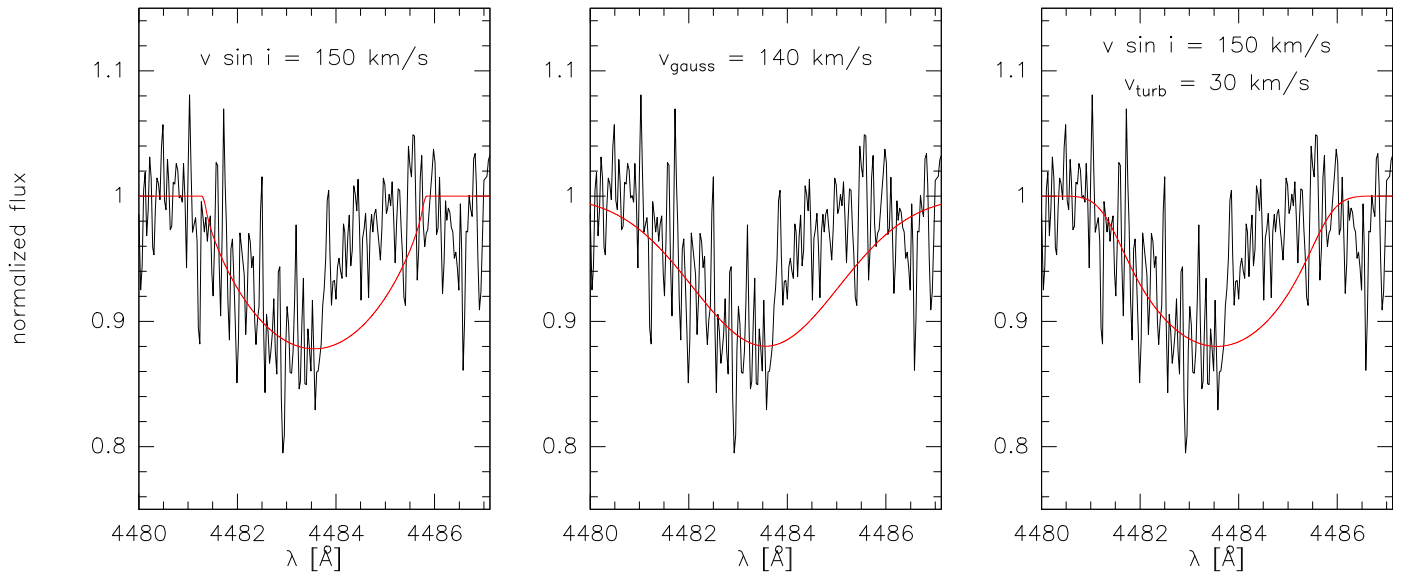}}
\caption{Fitting process of the Mg{\sc ii} $\lambda$4481 line with a pure
rotation profile (left panel), a pure Gaussian profile (mid panel), and
our best fit model (right panel) with a profile consisting mainly of stellar
rotation plus some macro-turbulence.}
\label{rotfit}
\end{figure*}
                                                                                 
To test which circumstellar extinction source might be acting on the $UBV$ 
fluxes of S\,23, we performed some simple test calculations for both scenarios.
For the circumstellar dust scenario, we adopted a grain size distribution
according to Mathis et al.\,(\cite{Mathis}) for either silicates or amorphous 
carbon grains
with SMC abundances of Si and C taken from Venn (\cite{Venn}). The optical 
depths of the circumstellar grains, resulting in a wavelength dependent dust 
extinction, are then calculated for the scenarios of an A0\,Ib and an A1\,Ib 
star. Interestingly, we found that neither the silicates nor the carbon grains 
(nor any combination of both) can account for the circumstellar extinction of 
S\,23. We therefore conclude that circumstellar dust is most probably not the 
(major) source of the extra extinction. The location of the circumstellar dust 
around S\,23, whose presence might be inferred from the IR photometry, should 
thus be out of the line of sight.

The situation is more complicated when the extinction is caused by the stellar
wind. In this case, the wind not only absorbs and scatters the stellar light,
instead it contributes with additional emission even at optical and UV
wavelengths, as has recently been shown by Kraus et al. (\cite{Kraus08}) for 
classical OB supergiants, and by Kraus et al. (\cite{Kraus07}) for the LMC B[e] 
supergiant \object{R\,126}.

We test the influence of the wind to the $UBV$ band fluxes by assuming for 
simplicity (and since nothing is known about the real structure and geometry of 
the wind around S\,23) a spherically symmetric isothermal wind, and calculate
the wind contribution with the model of Kraus et al. (\cite{Kraus08}). For the 
terminal wind velocity we use the escape velocity, which was found from 
observations of late B-type and early A-type supergiants by Lamers et al. 
(\cite{Lamers95}) to be a reasonable approximation. The only free 
parameter in the wind calculations is thus the mass loss rate, $\dot{M}$,
which is varied over a certain range to calculate the wind extinction
and emission in the $UBV$ bands.
 
Interestingly, we could not constrain the circumstellar extinction for the
scenario of an A0\,Ib star, while the scenario of an A1\,Ib star delivered 
reasonable values for both the mass loss rate, with $\dot{M} \simeq 3\times 
10^{-6}\,M_{\odot}$yr$^{-1}$, and the interstellar extinction, with $E(B-V) 
\simeq 0.03$, constraining the classification of S\,23 more towards an A1
star rather than an A0 star. 

Compared to the value found from the observed $B-V$ color index, our value is 
slightly lower, indicating that the neglect of the wind influence to the 
emergent flux from the atmosphere leads to an overestimation of the real 
interstellar extinction. Such behaviour has also recently been found for 
classical B-type supergiants (Kraus et al.\,\cite{Bsg}). 

With the tentative classification of S\,23 as A1\,Ib star with an effective
temperature of roughly 9000\,K, and with the derived interstellar extinction of 
$A_{V} \simeq 0.10$, we find the following set of stellar parameters: 
$\log (L_*/L_{\odot}) \simeq 4.3$, $R_{*} \simeq 55-60\,R_{\odot}$, and, from 
the comparison with stellar evolutionary tracks calculated for SMC stars by 
Charbonnel et al. (\cite{Charbonnel}), an initial mass of $M_{\rm ini} = 
9.5\ldots 11\,M_{\odot}$.

Finally, the small interstellar extinction value confirms the luminosity class 
adjustment of Ib, meaning that S\,23 belongs to the group of low-luminosity
B[e] supergiants, which has members in the LMC (Gummersbach et al.
\cite{Gummersbach}) as well as in the Milky Way (Borges Fernandes et al.
\cite{Marcelo}), and which should probably be extended at least to early
A-type stars.

\section{Stellar rotation}\label{rotation}

With the classification of S\,23 as an early A-type supergiant at the time of 
our observation, we next intend to determine its stellar rotation speed given 
by the parameter $\varv\sin i$. For late B-type and early A-type stars and
supergiants, the most reliable photospheric line for the stellar rotation 
determination is the Mg{\sc ii} $\lambda$4481 doublet (e.g., Royer \cite{Royer}; 
Royer et al. \cite{Royeretal}), which is also the only clearly detectable 
photospheric line in our spectra. 

It has recently been shown that the Fourier analysis provides an excellent 
tool for the rotation velocity determination even for hot stars (see Royer 
\cite{Royer}; Sim\'on-D\'iaz \& Herrero \cite{SimonDiaz}). However, for the 
Fourier analysis to work properly, it is necessary to have data with extremely 
good S/N ratio because otherwise the noise level is too high to allow the zero 
points caused by the rotation profile to show up in the Fourier transform of 
the spectral data (Sim\'on-D\'iaz \& Herrero \cite{SimonDiaz}). With the rather 
low S/N ratio of our FEROS data of the faint object S\,23, the powerful Fourier 
analysis is thus not applicable. Instead, we have to rely on the less accurate 
full-width at half-maximum ($FWHM$) method to extract the possible stellar 
rotation of the star, i.e., we fit 
synthetic line profiles to the observed, normalised Mg{\sc ii} line. The 
stellar rotation profile is described in the usual way (see, e.g., Gray 
\cite{Gray}) with a limb-darkening coefficient of $\epsilon = 0.6$.

The line profile of the Mg{\sc ii} $\lambda$4481 line appears to be asymmetric
(see Fig.\,\ref{rotfit}). However, this asymmetry is probably not real, 
but an artefact due to noise since only one of our spectra shows this 
asymmetry, while the other spectrum shows a symmetric line profile. For a 
proper $FWHM$ determination we therefore rely on the blue part of the line, 
only. Given the low S/N ratio, we derive a $FWHM$ of the observed Mg{\sc ii} 
$\lambda$4481 line of $3.5\pm 0.1\,$\AA. If we assume either a pure Gaussian 
profile or a pure rotation profile, we thus obtain maximum velocity 
values of $\varv_{\rm gauss} \simeq (140\pm 5)$\,km\,s$^{-1}$ and $\varv\sin i 
\simeq (150\pm 5)$\,km\,s$^{-1}$, respectively. However, both possibilities, 
i.e., either a pure Gaussian or a pure rotation profile, do not deliver 
satisfactory fits to the observed line profile (see left and middle panel of 
Fig.\,\ref{rotfit}). While the edges are too sharp in the case of pure stellar 
rotation, the wings of the pure Gaussian are way too broad. Instead, a 
combination of both might be a good solution. But how can we estimate the 
individual amounts of the Gaussian and the rotational components?

\begin{figure}[t!]
\resizebox{\hsize}{!}{\includegraphics{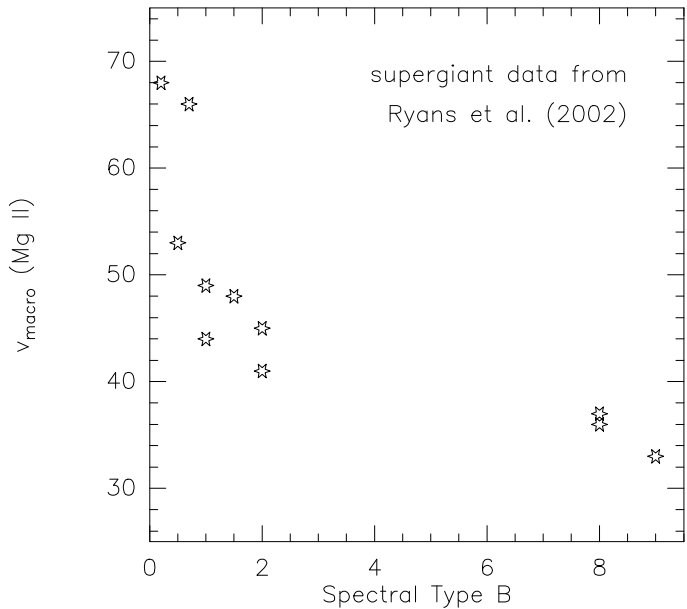}}
\caption{Maximum velocity caused by macro-turbulence to the Mg{\sc ii} 
$\lambda$4481 line for a sample of B-type supergiants. 
The data are taken from Ryans et al. (\cite{Ryans}).}
\label{vturb}
\end{figure}

It has been known for a long time that within the atmospheres of OBA 
supergiants some 
macroscopic line broadening occurs that is of no rotational origin. This 
broadening component can be described with a Gaussian profile and is usually
referred to as (macro-)turbulence (see, e.g., Howarth \cite{Howarth}). This
'extra' broadening of the photospheric lines is thereby not restricted to
hydrogen and helium lines, but also appears in metal lines. The
physical understanding and interpretation of this component is, however, still 
missing. In an attempt to determine the contribution of this macro-turbulence
to line broadening, Ryans et al. (\cite{Ryans}) investigated a sample of 
B-type supergiants trying to disentangle the contributions of rotation
and macro-turbulence with high precision. One of the key lines in their
investigation was the Mg{\sc ii} $\lambda$4481 line. Inspection of their
Table\,3 reveals that there is an obvious trend between the maximum possible 
turbulent velocity and the effective temperature of the star. This is shown in 
Fig.\,\ref{vturb}, where we plotted the maximum turbulent velocity
found for the sample supergiants of Ryans et al. (\cite{Ryans}) versus
spectral type of the stars. While for the early-type supergiants maximum 
macro-turbulence as high as 70\,km\,s$^{-1}$ might be present, it rapidly
decreases towards the late type stars to about 30--35\,km\,s$^{-1}$ only.
If this trend continues for the A-type supergiants, we might expect
a maximum macro-turbulence of about 30\,km\,s$^{-1}$ acting on the photospheric
lines of S\,23.

We use this value of 30\,km\,s$^{-1}$ for the Gaussian profile and calculate 
the Mg{\sc ii} $\lambda$4481 line profile by adjusting the rotational velocity 
such that the observed $FWHM$ is reproduced. Since the Gaussian velocity
contribution is rather small, it does not significantly influence the total 
line width, which is still purely rotationally dominated with about the same 
value of $\varv\sin i\sim 150\pm 5$\,km\,s$^{-1}$ as without macro-turbulence.
The only influence of the Gaussian component concerns the line wings (see
right panel of Fig.\,\ref{rotfit}). Overall, this combination of rotation
plus macro-turbulence gives a reasonably good fit to the observed line profile.

\begin{figure}[t!]
\resizebox{\hsize}{!}{\includegraphics{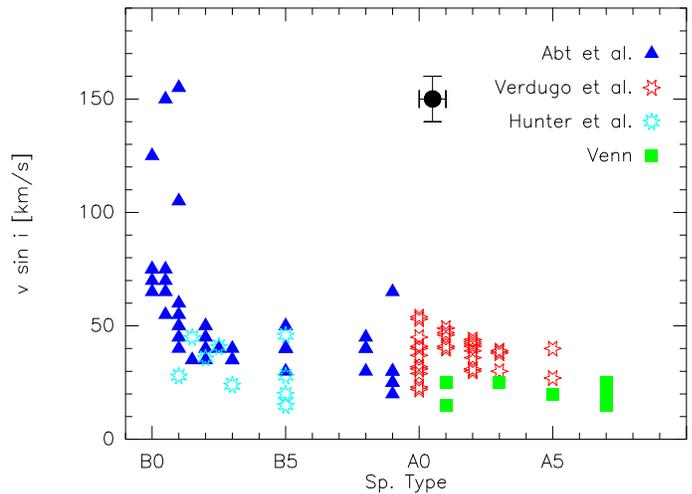}}
\caption{Distribution of projected rotational velocities with spectral type
of a sample of galactic (Abt et al. \cite{Abt}; Verdugo et al. \cite{Verdugo})
and SMC (Hunter et al. \cite{Hunter}; Venn\cite{Venn}) supergiants. The big 
spot with errorbars is the value of S\,23 derived from our analysis.}  
\label{rotvgl}
\end{figure}

The value of $\varv \sin i$ found from our fitting is high when compared
to the projected rotation velocities of normal galactic B and A-type 
supergiants, for which a tendency of decreasing rotation velocity with 
effective temperature is reported. This can be seen in Fig.\,\ref{rotvgl},
where we plot the results for four samples: (i) galactic B-type supergiants
from Abt et al. (\cite{Abt}); (ii) galactic A-type supergiants from Verdugo et 
al. (\cite{Verdugo}); (iii) SMC B-type supergiants from Hunter et al. 
(\cite{Hunter}); and (iv) SMC A-type supergiants from Venn (\cite{Venn}).
The SMC stars turn out to have even smaller velocities
than their galactic counterparts, but this effect might arise simply from
small number statistics. Compared to the mean rotational velocity of the
galactic and SMC late B-type and early A-type supergiants, the value derived 
for S\,23 is clearly a factor of three higher.

However, as discussed already in the introduction, the B[e] phenomenon seems to 
be linked to rapid stellar rotation. For the class of the B[e] supergiants, 
only for three stars the value of $\varv \sin i$ could be derived so far. 
These are the LMC B[e] supergiants \object{Hen\,S93} with $\varv \sin i \simeq
65$\,km\,s$^{-1}$ (Gummersbach et al.\,\cite{Gummersbach}) and \object{R\,66}
with $\varv \sin i \simeq 50$\,km\,s$^{-1}$ (Zickgraf \cite{Zickgraf2006}),
and the SMC B[e] supergiant \object{R\,50} with $\varv \sin i \simeq 
150$\,km\,s$^{-1}$ (Zickgraf \cite{Zickgraf2000}). The latter 
one has a projected rotation speed on the same order as S\,23, which is now
the fourth B[e] supergiant with confirmed high projected rotational velocity. 
From our derived set of stellar parameters we can further estimate the critical
velocity of the star, which turns out to be on the order of 200\,km\,s$^{-1}$.
This means, that S\,23 is rotating with at least 75\% of its critical velocity.

Interestingly, the optical spectrum of ZSW indicates that the Mg{\sc ii}
$\lambda$4481 line is less broad during the 1989 observations. A tentative 
derivation of its $FWHM$ delivers a value of ($2.6\pm 0.2$)\,\AA.~Neglecting
any other broadening mechanism, this $FWHM$ value results in a maximum 
rotational velocity of $\varv\sin i\simeq 110\pm 10$\,km\,s$^{-1}$. This 
result is surprising, because it indicates that S\,23 might not only have 
cooled, but at the same time also increased its projected rotational velocity 
by at least 35\%.

\section{Discussion}\label{discus}

The investigation of our high-resolution optical spectra in combination with
the photometric data of Massey (\cite{Massey}) resulted in the tentative
classification of S\,23 as a rapidly rotating A1\,Ib star. On the other hand, 
our evaluation of the Mg{\sc ii}/He{\sc i} line ratio derived (even though
with large error bars) from the spectra of ZSW, delivered a spectral 
classification more consistent with a star of roughly spectral type 
B9\,Ib. The most striking differences are thus an apparent temperature 
decrease by about 1500\,K, based on the spectral type versus effective 
temperature calibration of Evans \& Howarth (\cite{EvansHowarth}), and 
shown in Fig.\,\ref{ratio}; and, even more curious, an increase in projected 
rotational velocity by about 35\% (see previous section) within a time interval 
of only 11 years. In this section, we therefore investigate different scenarios 
that might explain this peculiar behaviour. For this, we first consider some 
natural possibilities like stellar evolution (Sect.\,\ref{stelevol}) and mass 
ejection occulting the star in form of a shell phase as is often observed for 
the classical Be stars (Sect.\,\ref{Be_comp}). Since both attempts fail to 
explain the observed peculiar variations in S\,23, we present and discuss in 
Sect.\,\ref{specul} further scenarios, which are more speculative. Of course, 
the amount and quality of our available data is definitely not sufficient for a 
proper analysis, which could lead to a discrimination between the different 
scenarios. Instead, we want to discuss qualitatively the expected observable 
effects for those scenarios that seem to be at least likely and worth further 
investigation.

\subsection{Stellar evolution}\label{stelevol}

The apparent cooling of the star in combination with a simultaneous speed-up
of the rotation velocity is unexpected and surprising. The first scenario 
we might think of is stellar evolution. However, stellar evolution theories 
predict that when the star expands, i.e., cools, it can only {\it decrease}
its rotation velocity. An {\it increase} in rotation speed during the post-main
sequence evolution of a rapidly rotating star at SMC metallicity and with an 
initial mass of about $10\,M_{\odot}$ can only occur during the blue loop, i.e.,
while the star contracts and heats up again (Maeder \& Meynet \cite{MaMe}).
In addition, the timescales for such a star to cool from about 11\,000\,K to 
less than 9\,500\,K is on the order of 5000\,yr (Meynet, private communication).
This is definitely longer than the one decade we found. The changes in the 
spectral appearance of S\,23 can thus definitely not be explained by current
theories of stellar evolution. 
 
\subsection{Comparison with shell formations of classical Be stars}\label{Be_comp}

A different interpretation of the apparent cooling might be given by the
assumption of material ejection, e.g., in form of a ring or shell, leading to 
an (partly) occultation of the central star. Such phenomenon are known, e.g., 
from Be stars that can undergo a so-called shell phase (see, e.g., Hanuschik
\cite{Hanuschik}; Porter \& Rivinius \cite{Porter}). The shell, ejected from
the central star, thereby acts as an extinction source.
Derivation of the stellar parameters from absorption lines created within
such a shell can, therefore, result in much lower {\it apparent} effective 
temperature values than the real stellar one derived during a non-shell
phase. Such a behaviour has been reported, e.g., for the Be shell star 4 
Herculis, for which an apparent drop in effective temperature by 3000\,K during
the shell phase was found by Koubsk\'y et al. (\cite{Koubsky}).
However, contrary to our target S\,23, but in agreement with expected physical
behaviour, the rotation velocity derived from the shell absorption line
decreased as well. A shell phase of S\,23, mimicking a temperature decrease, 
thus has severe problems in explaining our observations, even if the 
presence of some "shell-lines" (like the Fe{\sc ii} and Si{\sc ii} lines) 
in our spectra hint towards the existence of a large amount of circumstellar 
material.

\subsection{Possible (speculative) scenarios for S\,23}\label{specul}
 
Since neither stellar evolution, nor a high-density mass ejection in the form of
a ring or shell can explain the peculiarities observed for S\,23, we need to 
think of different mechanisms. In this section, we therefore present three 
alternative scenarios that might be worth further detailed investigation.
These are (i) a high precession rate, (ii) an eclipsing binary, and (iii) 
surface He abundance and/or temperature inhomogeneities. All three are able to 
explain the peculiarities of S\,23, and all three can be tested by future 
observations. But we want to stress again that these scenarios can only
be considered as simplified models, given the quality of the available data.

\subsubsection{Precession of the system}

One possible explanation for the apparent cooling with increasing rotation
velocity might be a precession of the star with respect to the line of sight. 
Since we know that S\,23 is rapidly rotating, the classification of the star
strongly depends on the inclination angle under which it is observed.

The effects of rapid stellar rotation are manifold. Rapid rotation leads to a 
deformation of the stellar surface in the form of flattening. Thus, the polar 
regions become much hotter than the equatorial region, which is usually 
referred to as gravity darkening. Depending on the fraction of the rotation 
velocity to the critical velocity, the difference between polar and equatorial 
effective temperature can exceed a factor of two. The same holds also for the
other stellar parameters like the surface gravity, stellar flux, 
mass flux, and escape velocity. 

From the differences in observed rotation velocity, $\varv\sin i$, we may 
conclude that the inclination of the system has increased between ZSW 
and our observations, which means, that the system was observed more edge-on
in 2000 than in 1989. To quantify this, we take the values of $\varv \sin i = 
110$\,km\,s$^{-1}$ in 1989, and $\varv\sin i=150$\,km\,s$^{-1}$ in 2000. 
Then, the difference in inclination follows to
\begin{equation}
\sin i_{\rm 2000} = \frac{150}{110} \sin i_{\rm 1989} = 1.36 \sin i_{\rm 1989},
\end{equation}
which means that the inclination angle has increased by almost 40\%. Assuming 
that the star was seen more or less edge-on (i.e., inclination angle of 
$90\degr$) in 2000, it must thus have been seen under an inclination angle of 
about $47\degr$ in 1989. This conclusion is, however, only strictly valid,
as long as the inclination did not cross the equatorial plane yet.

A lower limit to the inclination angle in 2000 is given by the critical 
rotation of the star, $\varv_{\rm crit} \simeq 200$\,km\,s$^{-1}$.
Assuming that the star is rotating at its critical speed, it delivers 
a minimum inclination angle of $48\degr$ in 2000, and of $35\degr$ in 1989.

\begin{figure}[t!]
\resizebox{\hsize}{!}{\includegraphics{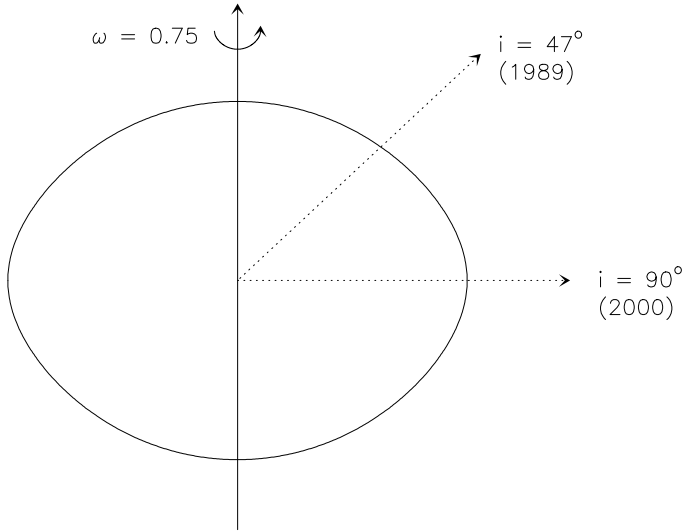}}
\caption{Shape of S\,23 rotating with 75\% of its critical velocity. Also 
included are the assumed positions of the observers with respect to the 
rotation axis of the star in the two periods of optical observations.}
\label{shape}
\end{figure}

\begin{figure}[t!]
\resizebox{\hsize}{!}{\includegraphics{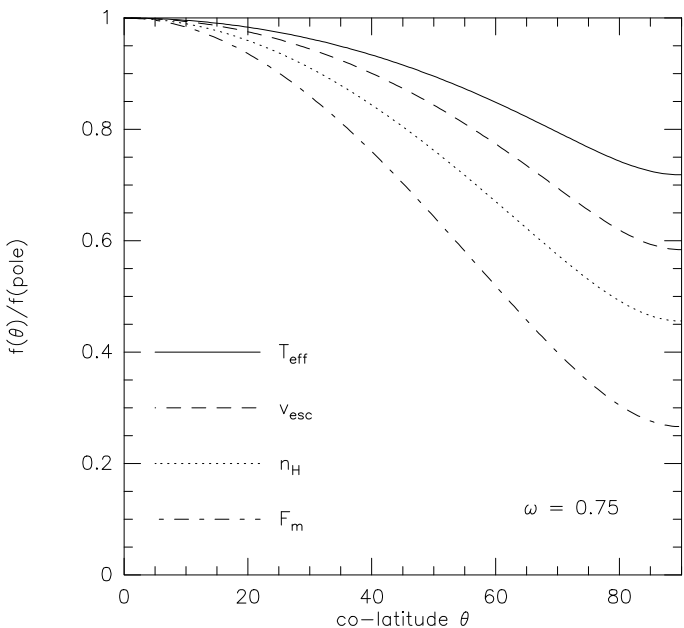}}
\caption{Distribution of the surface parameters (effective temperature, escape 
velocity, hydrogen density, and mass flux) with co-latitude, $\theta$, 
normalised to their polar values. 
All parameters are found to drop from pole to equator (for details on their
calculation see Kraus \cite{Kraus06}).}
\label{surface}
\end{figure}

To get a rough idea of the influence of the rapid rotation in combination
with the postulated precession on the observable mean effective temperature,
we consider the following scenario for S\,23 (which need not be correct, but 
which is only meant to show the basic idea of this toy model) : 

Since rotation at the critical limit is not very realistic, we assume in the
following that the star was seen more or less edge-on in 2000 so that S\,23 
rotates with 75\% of its critical speed, i.e., $\omega = \varv_{\rm eq}/
\varv_{\rm crit} = 0.75$. Such an edge-on scenario is not in contradiction 
with our results from Sect.\,\ref{extinc}, where we found that the dust is most 
probably not located along the line of sight. In fact, for rapidly rotating 
stars the mass loss happens much stronger via the poles than via the equatorial 
region, and to date models like, e.g., the wind-compressed disk (Bjorkman \& 
Cassinelli \cite{Bjorkman}) failed in confining the material within the 
equatorial plane. In addition, spectropolarimetric observations of Bjorkman et 
al. (\cite{Karen}) for the galactic B[e] star HD\,50138 have shown that the 
polarisation is mainly caused by scattering within an almost edge-on seen pure 
gaseous disk, while the huge amount of dust observed from this system seems not
to be connected with the disk at all.
                                                                                
Following Kraus (\cite{Kraus06}), we compute the surface structure of the star 
rotating with $\omega = 0.75$ (shown in Fig.\,\ref{shape}). For the adopted 
rotation speed, a deformation (i.e., flattening) of the stellar surface is 
obvious. Also included in this plot are the inclination angles under which the 
star might have been observed in 1989 and in 2000. While in 2000 most of the 
observed flux was emitted from the equatorial region with some minor 
contributions from higher altitudes, the flux observed in 1989 was emitted from 
an intermediate altitude, with contributions from both, the polar and the 
equatorial region.

To obtain an idea about the expected differences in the observed stellar 
spectra, we further calculate the surface distribution of the latitude 
dependent stellar parameters like effective temperature $T_{\rm eff}(\theta)$, 
escape velocity $\varv_{\rm esc}(\theta)$, mass flux $F_{\rm m}(\theta)$, and 
hydrogen desity $n_{\rm H}(\theta)$ for a star rotating with $\omega = 0.75$. 
These parameters are shown in Fig.\,\ref{surface} as a function of the 
co-latitude $\theta$ (i.e., $\theta = 0$ at the pole), normalised to their polar 
values. They are all found to decrease by non-negligible amounts from the
pole to the equator.

In particular, we find from Fig.\,\ref{surface} the following behaviour of the 
effective temperature:
\begin{eqnarray}
T_{\rm eff}(\theta = 47\degr) & \simeq & 0.91\, T_{\rm eff}(\theta = 0\degr)\, , \\
T_{\rm eff}(\theta = 47\degr) & \simeq & 1.26\, T_{\rm eff}(\theta = 90\degr)\, .
\end{eqnarray}
Since for an inclination of $\sim 47\degr$ the stellar emission contains 
contributions from the hotter as well as from the cooler regions, we might 
assume that the effective temperature at $\theta = 47\degr$ is reflecting
quite well the real observed spectrum. This means, that we can calculate
the polar as well as the equatorial value of the effective temperature,
resulting in $T_{\rm eff}(\theta = 0\degr) \simeq 12\,100$\,K, and
$T_{\rm eff}(\theta = 90\degr) \simeq 8\,700$\,K. The equatorial value is
somewhat lower than our effective temperature of 9\,000\,K derived from the 
FEROS spectrum. This is, however, no contradiction, because the real
``observable'' mean effective temperature under an edge-on inclination of the 
star will certainly be somewhat higher than the pure value at $90\degr$, 
due to some contributions from higher altitudes. In addition, the 
temperature calibration of Evans \& Howarth (\cite{EvansHowarth}) is certainly
too coarse for a detailed temperature assignment.

In summary, the precession scenario might explain both, the apparent cooling
with an increase in rotation velocity, just from an increase of the inclination
axis of a rapidly rotating star from about $47\degr$ to about $90\degr$.
Whether such a strong precession will indeed occur, and what might be the
reason for such rapid precession movement, is, however, not known.

To test the precession scenario, we would need to monitor the star over a
period of several years to see whether the He{\sc i} lines re-appear
(i.e., the star "heats up" again) with a simultaneous decrease of the projected
rotation velocity measured from the Mg{\sc ii} lines.

\subsubsection{An eclipsing binary scenario}

A second possible scenario to explain the strong variations observed in S\,23 
might be that of an eclipsing binary system, in which one component is 
perhaps a B9\,Ib star and the other one perhaps a A1\,Ib star. The 
disappearance of the He{\sc i} lines in our spectra might then be 
interpreted with the A1\,Ib star occulting more or less the hotter B9\,Ib 
component, while the spectra of ZSW, in which the He{\sc i} lines were
present and strong, indicate that the hotter component was dominating the 
spectral appearance of the system.

To test the hypothesis of S\,23 being an eclipsing binary, it
is certainly useful to perform a high S/N monitoring of its few available 
photospheric lines, (i.e., the He{\sc i} lines and the Mg{\sc ii} line). 
A further test for the eclipsing binary scenario would be the monitoring of its
light curve. Since the hotter component is also expected to be the more
luminous one, the system must appear fainter when the B9 star is eclipsed than
when the A1 star is eclipsed.

\subsubsection{A rapidly rotating star with surface inhomogeneities}

A third quite interesting scenario is based on the assumption that S\,23 might
have a patchy surface abundance in He. The existence of such inhomogeneous or
spotty surface patterns are known especially for some so-called chemically 
peculiar (CP) stars (see, e.g., Smith \cite{Smith}; Briquet et al.
\cite{Briquet01, Briquet04}; Lehmann et al. \cite{Lehmann06}; Krti\v{c}ka et 
al. \cite{Jirka}). These objects at or close to the main-sequence are supposed 
to be rigidly rotating, and the variations observed in their He{\sc i} line 
profiles are interpreted as abundance inhomogeneities in the form of large stellar
spots that appear within and again disappear from the line of sight as the star 
revolves. In addition, these stars are found to possess longitudinal magnetic 
fields (e.g., Khokhlova et al. \cite{Doppler}; Briquet et al. \cite{Briquet07}), 
and there seems to be evidence that variations in the magnetic field might be 
the cause for the abundance inhomogeneities.

Furthermore, the stellar surfaces of some of the CP stars seem to have
temperature inhomogeneities (see, e.g., B\"ohm-Vitense \& van Dyk \cite{BVvD}; 
Lehmann et al. \cite{Lehmann06}), and even the need of a co-existence of 
abundance and temperature inhomogeneities has recently been claimed by
Lehmann et al. (\cite{Lehmann07}), even though the physical processes 
resulting in such patchy surface structures are not understood at all.

S\,23 is definitely not a main-sequence star, and to our knowledge, no such 
abundance or temperature inhomogeneities have been reported for a supergiant
so far. In addition, nothing is known about a possible magnetic field on the
surface of S\,23. Nevertheless, such a surface inhomogeneity seems to be an 
interesting scenario, at least to explain the variations in the He{\sc i}
lines. Whether this inhomogeneity can also explain the different observed values of $\varv\sin
i$, needs to be investigated in more detail. At least the classical stellar 
rotation profile used in Sect.\,\ref{rotation} to derive $\varv\sin i$ 
from the $FWHM$ will in such a scenario certainly not be valid anymore.

If S\,23 possesses He spots, the variation of the He{\sc i} lines must be 
correlated to the rotation period of the star. Assuming that S\,23 is seen 
edge-on, the derived value of $\varv\sin i = 150$\,km\,s$^{-1}$ delivers the 
equatorial rotation velocity. With a radius of roughly 60\,$R_{\odot}$, this 
results in a period of about 20 days. To prove or disprove the scenario of 
surface inhomogeneities in either He abundance, or temperature, or even both, 
one would need to monitor the He{\sc i} and Mg{\sc ii} lines in the spectrum of 
S\,23 over at least one rotational period.

\section{Conclusions}\label{concl}

We investigated the SMC star S\,23, which was formerly classified by ZSW as a 
B8\,Ib star, belonging to the class of B[e] supergiants. Our optical spectra 
show clear differences to the previously published one. These differences 
especially concern the absence of all photospheric He{\sc i} absorption lines. 
Our spectra are therefore consistent more with a spectral classification of S\,23 
as A1\,Ib supergiant. This classification is further supported by our analysis 
of the photometric $UBV$ data. During this analysis, we could derive 
self-consistently the contributions of the interstellar extinction, and of the 
circumstellar extinction caused by the stellar wind. A circumstellar extinction
caused predominantly by dust could be excluded. 

We further determined the projected rotation velocity of S\,23 from an
analysis of its photospheric Mg{\sc ii} line. We found a value of
$\varv\sin i \simeq 150$\,km\,s$^{-1}$, which claims that S\,23 is rotating 
with at least 75\% of its critical velocity. Consequently, S\,23 is the fourth 
member of the B[e] supergiant star group with confirmed high projected
rotational velocity.

The most puzzling result is the fact that within a period of only 11 
years, between the two sets of spectroscopic observations, the star seems to
have cooled by more than 1500\,K, and at the same time increased its rotation
speed by about 35\%. Such behaviour is definitely ruled out by any stellar
evolution scenario. And also the scenario of a shell phase, as is often 
reported for classical Be stars, has severe difficulties in explaining all the
observed peculiarities in S\,23. Therefore, we presented and discussed several 
other alternatives: a rapidly rotating star undergoing a high precession, the 
scenario of an eclipsing binary, and the idea of He abundance and/or 
temperature inhomogeneities on the stellar surface, as they have been reported
from CP stars. All these suggestions can explain most of the curiosities
of S\,23. However, the question, which of these scenarios is the correct one, 
or whether even all scenarios have to be discarded, could not be solved. 
Nevertheless, we suggested observational tests that can be performed 
to distinguish or discard the different scenarios. Definitely more observations 
with a much better time resolution are needed to disentangle the 
mystery of the peculiar variations in S\,23.

%%%%%%%%%%%%%%%%%%%%%%%%%%%%%%%%%%%%%%%%%%%%%%%%%%%%%%%%%%%%%%%%%%%%%%%%%
                                                                                
\begin{acknowledgements}
                                                                                
We are greatful to Herman Hensberge and Georges Meynet for many fruitful 
discussions concerning the peculiarities of S\,23, and to the anonymous referee
for the helpful comments and suggestions.
This research made use of the NASA Astrophysics Data System (ADS) and of the
SIMBAD and 2MASS databases. M.K. acknowledges financial support from GA\,AV 
\v{C}R number KJB300030701, and M.K. and J.K. from GA\,\v{C}R number 
205/08/0003. M.B.F. acknowledges financial support from the Belgian Federal 
Science Policy Office (Research Fellowship for non-EU Postdocs) and CNRS for 
the post-doctoral grant.

\end{acknowledgements}
                                                                                
%%%%%%%%%%%%%%%%%%%%%%%%%%%%%%%%%%%%%%%%%%%%%%%%%%%%%%%%%%%%%%%%%%%%%%%%%

\end{document}